\documentclass[prd,aps,nofootinbib,floatfix,11pt]{revtex4}
\usepackage{amsmath,graphicx,epsfig,amssymb,dsfont,mathtools}
\usepackage{inputenc}
\usepackage[usenames]{color}
\usepackage{ulem} 
\usepackage{bigstrut}
\usepackage{slashed}
\usepackage{multirow}
\usepackage{subfigure}
\usepackage{makecell}

\allowdisplaybreaks

\begin{document}
\title{The study of the singly anti-charmed pentaquark production in b-factory}
\author{
  Xiao-Hui Hu~$^{1,2}$~\footnote{Email:huxiaohui@cumt.edu.cn}, Ye Xing~$^{1}$~\footnote{Corresponding author Email:xingye\_guang@cumt.edu.cn}}

\affiliation{$^{1}$ The College of Materials and physics, China University of mining and technology, Xuzhou 221116, China}
\affiliation{$^{2}$
Lanzhou Center for Theoretical Physics and Key Laboratory of Theoretical Physics of Gansu Province, Lanzhou University, Lanzhou 730000, China}
\begin{abstract}
  The b-factories, such as BelleII, BarBar and LHCb, emphasize the increasing importance of exotic hadron research. In this paper, we discuss the possible production of singly anti-charmed pentaquark states $\bar{c}q qqq$ from $B$ mesons in the b-factory under SU(3) symmetry analysis. Discussions of both possibilities have been driven by the hypothesis that the pentaquark state considered in this work, known as the lowest lying state $P_{\bar c}$, could be bound or unbound. We find the golden channels for the production of {the pentaquark ground states, such as $B^0 \to {P}_{\bar{c}sudu}^{(\prime) 0}\Sigma^0$}. We further estimate the branching ratios for the production of the ground states $P_{\bar c}$ from $B$ meson decays. Thus, multiple channels are available for experiments, which may remove certain obstacles to the discovery of new pentaquark states.
\end{abstract}
\maketitle

\section{Introduction}~\label{sec:Introduction}

The search for pentaquark states has emerged as a highly significant investigation within the quark model in recent years.
Theoretical and experimental achievements in this field have been outstanding in recent years~\cite{Chen:2016qju,Guo:2017jvc,Maiani:2023nwj,Wu:2010jy,Wang:2011rga,Karliner:2015ina,Chen:2015loa,Roca:2015dva,He:2015cea,Liu:2019zoy,Liu:2015fea,Mikhasenko:2015vca,Szczepaniak:2015hya,Esposito:2016noz,Zhu:2015bba,Li:2015gta,Lebed:2015tna,Ali:2020vee,An:2022fvs,LHCb:2015yax,LHCb:2019kea,LHCb:2020jpq,LHCb:2021chn}, particularly the b-factory experimental efforts. 
In 2015, the observations of $P_c(4380)^+$ and $P_c(4450)^+$~\cite{LHCb:2015yax} in the $\Lambda_{b}^{0}\to J/\psi pK^{-} $ decay were first announced by the LHCb collaboration. By analysing the $J/\psi p$ invariant mass distribution of $\Lambda_b^{0} \to K J/\psi p$, three pentaquark candidates $P_c(4440)^+$, $P_c(4457)^+$ and $P_c(4312)^+$ were found in 2019~\cite{LHCb:2019kea}. The split into two new overlapping peaks was then confirmed by analysis of the $\Xi_b^- \to J/\psi \Lambda K^-$ channel~\cite{LHCb:2020jpq}.
Using data collected from 2011 to 2018, LHCb performed an amplitude analysis of $B_s^0 \to J/\psi p \bar p$ in 2021 and found evidence for a new structure $P_{c}(4337)^+$~\cite{LHCb:2021chn}. This finding suggests that, in addition to the b-baryon system, $B$ decays could be valuable research areas for studying the pentaquark~\cite{Brodsky:1997yr,Gotzen:2005vm,Browder:2004mp, Drutskoy:2012gt}. The large number of accumulated $B$ mesons and their greater decay branching ratio may support the new route, which can simultaneously advance our probe research into the charmed pentaquark states. 
The b-factory, where a number of charmed hardrons have been discovered~\cite{LHCb:2022ine,HFLAV:2022pwe,Epifanov:2020elk,Eidelman:2020iul,ALICE:2021psx,ALICE:2020wla,ALICE:2020wfu,LHCb:2017iph,Rosales:2021qxn,BESIII:2013ris,LHCb:2015yax}, has been a discovery factory for exotic charmed hadrons. If the expected integrated luminosity of b-factories could reach around $300fb^{-1}\sim50ab^{-1}$, then $10^{10}\sim10^{13}$ pairs of $B^{0}\bar{B}^{0}$($B^{+}{B}^{-}$/$B^{0}_{s}\bar{B}^{0}_{s}$) will be produced in the future~\cite{Wang:2022nrm,Belle-II:2018jsg,LHCb:2018roe,LHCb:2016qpe,LHCb:2019fns,HFLAV:2019otj,LHCb:2019tea}.  
Therefore, with the continuous improvement of experimental detection capability, more charmed pentaquark states will be discovered by b-factory experiments in the future.

To date, the majority of $P_c$ states, also known as hidden charmed pentaquarks, have been found in the $J/\psi p$ invariant mass distribution. Their thresholds are determined by a baryon and a meson close to the mass of the $P_c$ states. For example, the threshold of $\Sigma_{c}\bar{D}$ is close to the pentaquark $P_c(4312)^+$, and that of $\Sigma_{c}\bar{D}^{*}$ is close to the states $P_c(4450)^+$ and $P_c(4457)^+$. For the five-quark system, its diversified composition will depend on different production mechanisms. Assuming that the pentaquark state is compact, it can be bound by gluon exchange~\cite{Ali:2020vee,Ali:2019clg}.
Fortunately, the mass spectrum and decay properties~\cite{Wu:2017weo,Takeuchi:2016ejt} indicate that this compact configuration is not incompatible with the observed $P_c$ state.
If several particles interact appropriately with each other, they can form a weakly bound molecular state. To produce the loose molecules, the energetic quarks can first fragment into multiple hadrons, and then the strong interactions between these hadrons can produce some possible hadronic molecules. Therefore, the production of pentaquark states can be studied at the hadron level~\cite{Jin:2016vjn}.
However, the inner structure has been credibly described by the high-energy experiments, which may be far more complex than those of conventional hadrons, and the spin-parity is still unknown.
There have been several theoretical approaches to describe the pentaquark states, including quark models~\cite{Santopinto:2016pkp,Deng:2016rus,Ali:2019clg,Maiani:2018tfe,Stancu:1998sm}, meson-based model~\cite{Du:2019pij,Wang:2019ato,Chen:2015loa}, hadro-charmonium model~\cite{Dubynskiy:2008mq,Li:2013ssa,Wang:2013kra}, QCD sum rules~\cite{Chen:2012ex,Chen:2014vha}, kinematical effects~\cite{Guo:2015umn}, chiral quark model~\cite{Qin:2020gxr}, heavy quark symmetry~\cite{Ali:2016dkf}, complex scaling method~\cite{Lin:2023dbp} and more. By assuming that the pentaquark state is a bound state of two diquarks and an anti-quark, the authors calculate the masses of the pentaquark ground state~\cite{abu-shady:2022apq}. They used the Bethe-Salpeter equation, with the potential energy of quark interaction given by logarithm, linear, and spin dependent potentials.

In addition, the open charmed pentaquark state will provide a new opportunity for the current study of the charmed pentaquark system. 
Based on the difference of charmed quark, the open charmed pentaquark state can be divied into two kinds: anti-charmed pentaquark state ($\bar{c}qqqq$) and charmed pentaquark state ($c\bar{q}qqq$).
In this work, we will focus on the anti-charmed pentaquark states, 
which have already been discussed by different approaches~\cite{Diakonov:2010tf,Yamaguchi:2011xb,Richard:2019fms,Giannuzzi:2019esi,An:2020vku,Stewart:2004pd,Chow:1995hv,Leung:1998tt,Yang:2020atz,Jaffe:2003sg,Sarac:2005fn,Azizi:2021pbh}. 
And the descriptions of the inner structure from different approaches disagree with each other, so need more researches on them. Accordingly, we will force the production of the singly anti-charmed pentaquark, predicting the adopted channels for future experiments. In return, the discussion may help us understand the interactions between inner particles and the confinement mechanism.
In this work, we will discuss the production of anti-charmed pentaquark states via $B$ meson decays at the b-factories, using the SU(3) symmetry analysis approach. 
The flavor SU(3) symmetry~\cite{Wang:2019dls,Shi:2017dto,Xing:2019hjg,Li:2021rfj} is a convincing tool to analyze the production and decay behaviors of hadrons.

The rest of the paper is organized as follows. In Sec.~\ref{sec:Pentaquark}, we represent the tensor representations of anti-charmed pentaquark states and argue whether the ground states 
exceed the threshold for strong decay. In Sec.~\ref{sec:Decays}, we debate both the strong and weak decays of the pentaquark ground states. In Sec.~\ref{sec:production}, we provide the potential production Hamiltonian of the pentaquark ground states from $B$ meson decays and select out the golden production channels for ground states. We examine the production of the ground states from bottomed mesons and baryons at the b-factory. In Sec.~\ref{sec:summary}, a short summary is given. In Sec.~\ref{sec:appendix}, the production of the pentaquark exotic states $\boldsymbol{15}$ and $\boldsymbol{15^{\prime}}$ from $B$ meson decays is shown in detail.

\section{Pentaquark $\bar c qqqq$}~\label{sec:Pentaquark}
In this section, we will mainly discuss the representation of singly anti-charmed pentaquark state with $1/2^-$ spin-parity. {The singly anti-charmed pentaquark states $\bar{c}qqqq$ can be decomposed multiplets: $\boldsymbol{3}$, $\boldsymbol{\bar6}$, $\boldsymbol{15}$ and $\boldsymbol{15^{\prime}}$.
The exotic states $\boldsymbol{15}$ and $\boldsymbol{15^{\prime}}$ can strongly decay into ground states $\boldsymbol{3}$ and $\boldsymbol{\bar6}$, so the focus of this work is on the anti-charmed pentaquark ground states $\boldsymbol{3}$ and $\boldsymbol{\bar6}$, noted with $P_{\bar{c}\boldsymbol{3}}$ and $P_{\bar{c}\boldsymbol{\bar6}}$.}
Their representation in \textbf{flavor} $\boldsymbol{\otimes}$ \textbf{color} $\boldsymbol{\otimes}$ \textbf{spin} space satisfying Fermi statistics, should be global antisymmetry. 

{In flavor space, 
the matrix of the pentaquark ground states $\bar{c}\big[[qq]_{\boldsymbol{\bar 3}}[qq]_{\boldsymbol{\bar 3}}\big]_{\boldsymbol{3}}$ and $\bar{c}\big\{[qq]_{\boldsymbol{\bar 3}}[qq]_{\boldsymbol{\bar 3}}\big\}_{\boldsymbol{\bar 6}}$ can be given as following,
  \begin{eqnarray}
    && (P_{\bar{c}\boldsymbol{3}})=\left(\begin{array}{c}
      \bar{c}[[su][du]]\\
      \bar{c}[[ds][du]]\\
      \bar{c}[[ds][su]]\\
  \end{array}\right)=\left(\begin{array}{c}
      P_{\bar{c}sudu}^{0}\\
      P_{\bar{c}dsdu}^{-}\\
      P_{\bar{c}dssu}^{-}\\
  \end{array}\right),\\
     && (P_{\bar{c}\boldsymbol{\bar6}})=\left(\begin{array}{ccc}
          \bar{c}\{[ds][ds]\}&\bar{c}\{[ds][su]\}&\bar{c}\{[ds][du]\}\\
          \bar{c}\{[ds][su]\}&\bar{c}\{[su][su]\}&\bar{c}\{[su][du]\}\\
          \bar{c}\{[ds][du]\}&\bar{c}\{[su][du]\}&\bar{c}\{[du][du]\}\\
      \end{array}\right)=\left(\begin{array}{ccc}
          P_{\bar{c}dsds}^{--}&{P}_{\bar{c}dssu}^{\prime-}&{P}_{\bar{c}dsdu}^{\prime-}\\
          {P}_{\bar{c}dssu}^{\prime-}&{P}_{\bar{c}susu}^{0}&{P}_{\bar{c}sudu}^{\prime 0}\\
          {P}_{\bar{c}dsdu}^{\prime-}&{P}_{\bar{c}sudu}^{\prime 0}&{P}_{\bar{c}dudu}^{0}\\
      \end{array}\right),
  \end{eqnarray}
here the curly braces and square brackets indicate the symmetry (S) and antisymmetry (A) of the four light quarks respectively. The superscript of the pentaquark state $P$ denotes the charge, and the index represents the flavor constituent.}

According to the good diquark scheme~\cite{Jaffe:2004ph}, in color space the good diquark should be antisymmetry as $[qq]_{\boldsymbol{\bar 3}}(A)$, 
and in spin space the good diquark is a spinless boson, making it antisymmetric as $[qq]\to 0_{s}(A)$.
Thus, under \textbf{flavor} $\boldsymbol{\otimes}$ \textbf{color} $\boldsymbol{\otimes}$ \textbf{spin} space, the wave function of the singly anti-charmed pentaquark ground states $\boldsymbol{3}$ and $\boldsymbol{\bar 6}$ can be written roughly as follows:
  \begin{eqnarray}
  \psi_{\text{flavor}}=
  \left\{
    \begin{array}{c}
      \bar{c}\Big[[qq]_{\boldsymbol{\bar{3}}}[qq]_{\boldsymbol{\bar{3}}}\Big]_{\boldsymbol 3}\\ 
      \bar{c}\Big\{[qq]_{\boldsymbol{\bar{3}}}[qq]_{\boldsymbol{\bar{3}}}\Big\}_{\boldsymbol{\bar 6}}
    \end{array}\right.,\ \
  \psi_{\text{color}}=
  \bar{c}_{\boldsymbol{\bar{3}}}\Big [[qq]_{\boldsymbol{\bar{3}}}[qq]_{\boldsymbol{\bar{3}}}\Big]_{3},\ \
  \psi_{\text{spin}}=
  \left\{
    \begin{array}{c}
      \bar{c}_{1\over 2}\Big[[qq]_{0}[qq]_{0}\Big]_{0}\\
  \bar{c}_{1\over 2}\Big\{[qq]_{0}[qq]_{0}\Big\}_{0}
\end{array}\right..
  \end{eqnarray}
  
The masses of the ground states have been calculated by various methods~\cite{Jaffe:2003sg,Stewart:2004pd,Richard:2019fms,An:2020vku}. Several representative results are listed in Tab.~\ref{tab:mass}, and among them, those predicted by QCD sum rules~\cite{Sarac:2005fn} are all lower than $3~\rm{GeV}$.
If the pentaquark ground states are below their respective strong decay thresholds, they will decay weakly. Otherwise, they will decay strongly. In the chromomagnetic interaction model (CIM), {${P}_{\bar{c}dudu}^{0}$} is above the strong decay threshold $DN$ by about $66\ \rm{MeV}$, but the quark model~(QM) shows the opposite conclusion, with {${P}_{\bar{c}dudu}^{0}$} below the threshold by about $94\ \rm{MeV}$. The same controversy exists in the cases of {$P_{\bar{c}sudu}^{(\prime) 0}$}, {${P}_{\bar{c}dsdu}^{(\prime)-}$} and {${P}_{\bar{c}dssu}^{(\prime)-}$}. In the simple quark model and the constituent quark model, {$P_{\bar{c}sudu}^{(\prime) 0}$} and {${P}_{\bar{c}dsdu}^{(\prime)-}$} are believed to be stable and all below the threshold $Dp$ of about $326\ \rm{MeV}$ and $75\ \rm{MeV}$. In contrast, it is slightly higher than their threshold of about 52 MeV in the CIM method. Moreover, the mass of {${P}_{\bar{c}dssu}^{(\prime)-}$} should be above the strong decay threshold $D_s\Lambda$ of $32\ \rm{MeV}$ in the constituent model, but below the threshold in the simple quark model and the CIM method. It is worth noting that the QCD sum rules predict that all pentaquark ground states can be lower than $3\ \rm{GeV}$, which contradicts the CIM method in the discussion of the {${P}_{\bar{c}susu}^{0}$} and {$P_{\bar{c}dsds}^{--}$} states.
In this work, we would fully consider all possible decay modes of the pentaquark ground states. 
\begin{table}
\caption{The masses of the pentaquark ground states.}\label{tab:mass}
\begin{tabular}{c|c |c |c| c| c|c }\hline\hline
mass/\rm{GeV}&{$P_{\bar{c}sudu}^{(\prime) 0}$}&{${P}_{\bar{c}dsdu}^{(\prime)-}$}&{${P}_{\bar{c}dssu}^{(\prime)-}$}&{${P}_{\bar{c}dudu}^{0}$}&{${P}_{\bar{c}susu}^{0}$}&{$P_{\bar{c}dsds}^{--}$}\\\hline
Quark model&2.580\cite{Stewart:2004pd}&2.580\cite{Stewart:2004pd}&2.77\cite{Stewart:2004pd}&2.71~\cite{Jaffe:2003sg}&-&-\\\hline
Constituent model~\cite{Stancu:1998sm,Richard:2019fms}&2.958~\cite{Richard:2019fms}&2.958~\cite{Richard:2019fms}&3.116~\cite{Richard:2019fms}&2.895~\cite{Stancu:1998sm}&-&-\\\hline
Chromomagnetic Interaction model~\cite{An:2020vku}&2.831&2.831&3.026&2.87&3.22&3.22\\\hline
QCD sum rules~\cite{Sarac:2005fn}&\multicolumn{6}{c}{$<3.0$}\\\hline\hline
\end{tabular}
\end{table}

\section{Decays of the ground states}~\label{sec:Decays}
In this section, we will discuss in detail the strong and weak decays of the pentaquark ground states.
The ground states  can be unbound states, which are unstable and decay immediately through the strong interaction. The Hamiltonian of the decays at the hadronic level can be directly driven as follows:
{
\begin{eqnarray}
  &&{\cal{H}}_{\text{s}}
  =a_{1}(P_{\bar{c}\boldsymbol{3}})_{i}(D)_{j}(P_{8})^{\{ij\}}
  +A_{1}(\tilde{P}_{\bar{c}\boldsymbol{\bar{6}}})_{\{\alpha k\}}(D)_{i}(P_{8})_{j}^{k}\varepsilon^{\alpha ij},
\end{eqnarray}
where the $a_{1}$ ($A_{1}$) term represents the pentaquark ground states $\boldsymbol{3} (\boldsymbol{\bar{6}})$ decaying into the $D$ meson and the light baryon.}
Light baryons consist of three light quarks, which can form an SU(3) octet $P_8$~\cite{Shi:2017dto,Li:2021rfj}. In SU(3) flavor space, the octet has the following matrix,
\begin{eqnarray}
P_8= \left(\begin{array}{ccc} \frac{1}{\sqrt{2}}\Sigma^0+\frac{1}{\sqrt{6}}\Lambda & \Sigma^+  &  p  \\ \Sigma^-  &  -\frac{1}{\sqrt{2}}\Sigma^0+\frac{1}{\sqrt{6}}\Lambda & n \\ \Xi^-   & \Xi^0  & -\sqrt{\frac{2}{3}}\Lambda
  \end{array} \right) \,.
\end{eqnarray}
The expansion of the Hamiltonian can lead to amplitudes for the strong decays of unbound states. 
{Considering the detection efficiency, some channels with less important contributions can be excluded from Tab.~\ref{tab:strongdecay}. We will focus on final states that are most likely to be detected in experiments and disregard those that are difficult to detect.} 
All channels with the hadrons $\pi^0, n, \Sigma^+(\to p\pi^0), \Sigma^-(\to n\pi^-)$, $\Xi^0(\to \Lambda \pi^0)$ in the final states are removed. And the processes with $\pi^{\pm}, \Sigma^0(\to N\pi\gamma)$, $\Xi^-(\to \Lambda \pi^-)$ and $\Lambda(\to p\pi^-)$ are kept.
{
\begin{eqnarray}
  \label{eq:Pc6goldenstrongdecay}
&&{P}_{\bar{c}dudu}^{0}\to  D^- p, \ {P}_{\bar{c}sudu}^{(\prime) 0} \to  D_s^- p,\ ({P}_{\bar{c}sudu}^{(\prime) 0}\to   \overline D^0  \Lambda,\ {P}_{\bar{c}sudu}^{(\prime) 0}\to   \overline D^0  \Sigma^0),\nonumber\\
&&P_{\bar{c}dsds}^{--}\to   D^-  \Xi^-, \ {P}_{\bar{c}dsdu}^{(\prime)-}\to  D^- \Lambda,\ ( {P}_{\bar{c}dsdu}^{(\prime)-} \to D^- \Sigma^0), \nonumber\\
&&
{P}_{\bar{c}dssu}^{\prime-}\to    D^-_s  \Sigma^0,\ {P}_{\bar{c}dssu}^{(\prime)-}\to   \overline D^0  \Xi^-, P_{\bar{c}dssu}^{-}\to D^-_s  \Lambda.
\end{eqnarray}}
Whereas $2\sim 3$ channels are prepared for the experimental reconstruction of {$P_{\bar{c}sudu}^{(\prime) 0}$}, {${P}_{\bar{c}dsdu}^{(\prime)-}$} and {${P}_{\bar{c}dssu}^{(\prime)-}$}.
We consider the strong decays of the ground states {${P}_{\bar{c}dudu}^{0}$, $P_{\bar{c}sudu}^{(\prime) 0}$, ${P}_{\bar{c}dsdu}^{(\prime)-}$, $P_{\bar{c}dsds}^{--}$ and ${P}_{\bar{c}dssu}^{(\prime)-}$}. Once the amplitudes of the different modes are obtained, ignoring the phase space effect, the relationships between the different decay channels are immediately deduced.
{
\begin{eqnarray}
&& \Gamma(P_{\bar{c}sudu}^{0}\to  D^-_s p)=6\Gamma(P_{\bar{c}sudu}^{0}\to \overline D^0  \Lambda)=2\Gamma(P_{\bar{c}sudu}^{0}\to   \overline D^0  \Sigma^0)= 6\Gamma(P_{\bar{c}dsdu}^{-}\to D^- \Lambda)\nonumber\\
&&= 2\Gamma(P_{\bar{c}dsdu}^{-}\to D^- \Sigma^0)= \Gamma(P_{\bar{c}dssu}^{-}\to \overline{D}^0 \Xi^-)=\frac{3}{2}\Gamma(P_{\bar{c}dssu}^{-}\to  D^-_s \Lambda^0),\label{eq:resd3}\\
&& \Gamma({P}_{\bar{c}dudu}^{0}\to D^- p)=\Gamma({P}_{\bar{c}sudu}^{\prime 0}\to  D^-_s p)= \frac{2}{3}\Gamma({P}_{\bar{c}dsdu}^{\prime-}\to D^- \Lambda)\nonumber\\
&&=\frac{2}{3}\Gamma({P}_{\bar{c}sudu}^{\prime 0}\to \overline D^0  \Lambda)=2\Gamma({P}_{\bar{c}sudu}^{\prime 0}\to   \overline D^0  \Sigma^0)=\Gamma(P_{\bar{c}dsds}^{--}\to   D^-  \Xi^-)\nonumber\\
&&= 2\Gamma({P}_{\bar{c}dsdu}^{\prime-}\to D^- \Sigma^0)= \Gamma({P}_{\bar{c}dssu}^{\prime-}\to \overline{D}^0 \Xi^-)=\frac{1}{2}\Gamma({P}_{\bar{c}dssu}^{\prime-}\to  D^-_s \Sigma^0).\label{eq:resd6}
  \end{eqnarray}}

 \begin{table}
  \caption{{The amplitude for the strong decays of the pentaquark ground states {$P_{\bar{c}\boldsymbol{3}}$ and $P_{\bar{c}\boldsymbol{\bar 6}}$} into a D meson and one light baryon.}}\label{tab:strongdecay}
  \begin{tabular}{|lc|lc|lc|}\hline\hline
    channel & amplitude &channel & amplitude&channel & amplitude\\\hline
  $P_{\bar csudu}^{0}\to   \overline D^0  \Lambda^0 $ & $ \frac{a_1}{\sqrt{6}}$&
  $P_{\bar csudu}^{0}\to   \overline D^0  \Sigma^0 $ & $ \frac{a_1}{\sqrt{2}}$&
  $P_{\bar csudu}^{0}\to   D^-  \Sigma^+ $ & $ a_1$\\
  $P_{\bar csudu}^{0}\to    D^-_s  p $ & $ a_1$&
  $P_{\bar cdsdu}^{-}\to   \overline D^0  \Sigma^- $ & $ a_1$&
  $P_{\bar cdsdu}^{-}\to   D^-  \Lambda^0 $ & $ \frac{a_1}{\sqrt{6}}$\\
  $P_{\bar cdsdu}^{-}\to   D^-  \Sigma^0 $ & $ -\frac{a_1}{\sqrt{2}}$&
  $P_{\bar cdsdu}^{-}\to    D^-_s  n $ & $ a_1$&
  $P_{\bar cdssu}^{-}\to   \overline D^0  \Xi^- $ & $ a_1$\\
  $P_{\bar cdssu}^{-}\to   D^-  \Xi^0 $ & $ a_1$&
  $P_{\bar cdssu}^{-}\to    D^-_s  \Lambda^0 $ & $ -\sqrt{\frac{2}{3}} a_1$&&\\\hline
  ${P}_{\bar cdudu}^{0}\to   \overline D^0  n $ & $ A_1$&
  ${P}_{\bar cdudu}^{0}\to   D^-  p $ & $ -A_1$&
  ${P}_{\bar csudu}^{\prime 0}\to   \overline D^0  \Lambda^0 $ & $ \sqrt{\frac{3}{2}} A_1$\\
  ${P}_{\bar csudu}^{\prime 0}\to   \overline D^0  \Sigma^0 $ & $ -\frac{A_1}{\sqrt{2}}$&
  ${P}_{\bar csudu}^{\prime 0}\to   D^-  \Sigma^+ $ & $ -A_1$&
  ${P}_{\bar csudu}^{\prime 0}\to    D^-_s  p $ & $ A_1$\\
  ${P}_{\bar cdsdu}^{\prime-}\to   \overline D^0  \Sigma^- $ & $ A_1$&
  ${P}_{\bar cdsdu}^{\prime-}\to   D^-  \Lambda^0 $ & $ -\sqrt{\frac{3}{2}} A_1$&
  ${P}_{\bar cdsdu}^{\prime-}\to   D^-  \Sigma^0 $ & $ -\frac{A_1}{\sqrt{2}}$\\
  ${P}_{\bar cdsdu}^{\prime-}\to    D^-_s  n $ & $ -A_1$&
  ${P}_{\bar csusu}^{0}\to   \overline D^0  \Xi^0 $ & $ -A_1$&
  ${P}_{\bar csusu}^{0}\to    D^-_s  \Sigma^+ $ & $ A_1$\\
  ${P}_{\bar cdssu}^{\prime-}\to   \overline D^0  \Xi^- $ & $ -A_1$&
  ${P}_{\bar cdssu}^{\prime-}\to   D^-  \Xi^0 $ & $ A_1$&
  ${P}_{\bar cdssu}^{\prime-}\to    D^-_s  \Sigma^0 $ & $ \sqrt{2} A_1$\\
  ${P}_{\bar cdsds}^{--}\to   D^-  \Xi^- $ & $ A_1$&
  ${P}_{\bar cdsds}^{--}\to    D^-_s  \Sigma^- $ & $ -A_1$&&\\\hline
    \hline
    \end{tabular}
    \end{table}

The stable pentaquark states are widely accepted and are dominated by weak decays. At the quark level, the Cabibbo allowed transition should be $\bar c\to \bar s d \bar u$.
Thus, the Hamiltonian for the weak decay of the ground states {$P_{\bar{c}\boldsymbol{3}}$ and $P_{\bar{c}\boldsymbol{\bar 6}}$} can be constructed as follows:
{
\begin{eqnarray}
  &&{\cal{H}}_{\text{w}}={b}_1 (P_{\bar{c}\boldsymbol{3}})_{i}(H_{\boldsymbol{6}})^{[ij]}_{k} M_{l}^{k} (P_{8})_{j}^{l}+{b}_2 (P_{\bar{c}\boldsymbol{3}})_{i}(H_{\boldsymbol{6}})^{[ij]}_{l} M_{j}^{k} (P_{8})_{k}^{l}\nonumber\\
  &&\quad +{b}_3 (P_{\bar{c}\boldsymbol{3}})_{k}(H_{\boldsymbol{6}})^{[ij]}_{l} M_{i}^{k} (P_{8})_{j}^{l}+{b}_4 (P_{\bar{c}\boldsymbol{3}})_{l}(H_{\boldsymbol{6}})^{[ij]}_{k} M_{i}^{k} (P_{8})_{j}^{l}\nonumber\\
  &&\quad +{B}_1  ({P}_{\bar{c}\boldsymbol{\bar 6}})^{\{\alpha i\}} (H_{\boldsymbol{6}})^{[jk]}_i M_{j}^{l} (P_{8})_{k}^{m} \varepsilon_{\alpha lm}+{B}_2 ({P}_{\bar{c}\boldsymbol{\bar 6}})^{\{\alpha i\}} (H_{\boldsymbol{6}})^{[jk]}_i M_{j}^{l} (P_{8})_{l}^{m}\varepsilon_{\alpha km}\nonumber\\
    &&\quad +{B}_3 ({P}_{\bar{c}\boldsymbol{\bar 6}})^{\{\alpha i\}} (H_{\boldsymbol{6}})^{[jk]}_i M_{m}^{l} (P_{8})_{j}^{m}\varepsilon_{\alpha kl}+{B}_4 ({P}_{\bar{c}\boldsymbol{\bar 6}})^{\{\alpha i\}} (H_{\boldsymbol{6}})^{[jk]}_l M_{i}^{l} (P_{8})_{j}^{m}\varepsilon_{\alpha km} 
    \nonumber\\
    &&\quad +{B}_5  ({P}_{\bar{c}\boldsymbol{\bar 6}})^{\{\alpha i\}} (H_{\boldsymbol{6}})^{[jk]}_l M_{j}^{l} (P_{8})_{i}^{m}\varepsilon_{\alpha km}+{B}_6  ({P}_{\bar{c}\boldsymbol{\bar 6}})^{\{\alpha i\}} (H_{\boldsymbol{6}})^{[jk]}_m M_{i}^{l} (P_{8})_{j}^{m}\varepsilon_{\alpha kl} \nonumber\\
    &&\quad +{B}_7  ({P}_{\bar{c}\boldsymbol{\bar 6}})^{\{\alpha i\}} (H_{\boldsymbol{6}})^{[jk]}_m M_{j}^{l} (P_{8})_{i}^{m} \varepsilon_{\alpha kl}+H_{\boldsymbol{6}}\to H_{15}.
    \label{eq:weakhami}
  \end{eqnarray}
The transition operator $\bar c \to \bar{s} d \bar{u}$ can be composed as $\boldsymbol{3\otimes 3\otimes \bar 3}=\boldsymbol{\bar 3\oplus \bar3\oplus 6\oplus 15}$. Because there is no penguin diagram of the transition $\bar c \to \bar{s} d \bar{u}$, the contribution of $H_{\boldsymbol{\bar 3}}$ will be vanished. Then we only consider the other two Cabibbo allowed non-zero tensor components, $(H_{\boldsymbol{6}})^{31}_2=-(H_{\boldsymbol{6}})^{13}_2=1$ and $(H_{\boldsymbol{15}})^{31}_2=(H_{ \boldsymbol{15}})^{13}_2=1$.}
Based on the Hamiltonian, the decay amplitude can be calculated and given in Tab.~\ref{tab:Pc6_weakdecay}. We choose some golden channels to help the experiment find the ground states.
{
\begin{eqnarray}\label{eq:Pc6goldenweakdecay}
&&{P}_{\bar{c}sudu}^{(\prime) 0}\to   \pi^-   p ,\ {P}_{\bar{c}susu}^{0}\to   K^-   p,\ ({P}_{\bar{c}susu}^{0}\to   K^+   \Xi^-), \ {P}_{\bar{c}dssu}^{(\prime)-}\to   \pi^-   \Lambda,\ ({P}_{\bar{c}dssu}^{(\prime)-}\to   \pi^-   \Sigma^0).
\end{eqnarray}}
Accordingly, the states {{${P}_{\bar{c}susu}^{0}$}}, {$P_{\bar{c}sudu}^{(\prime) 0}$} and {${P}_{\bar{c}dssu}^{(\prime)-}$} can occupy the largest potential in the experimental search. Therefore, in this work, we would force the weak decays of the states {{${P}_{\bar{c}susu}^{0}$}}, {$P_{\bar{c}sudu}^{(\prime) 0}$} and {${P}_{\bar{c}dssu}^{(\prime)-}$}. In addition, a candidate channel has been prepared for the reconstruction of {${P}_{\bar{c}susu}^{0}$} or {${P}_{\bar{c}dssu}^{(\prime)-}$} in brackets.

 \begin{table}
 \caption{{The Cabibbo allowed channels for the ground states to decay into light mesons and baryons states. The parameters $\bar{b}_{i}$ and $\bar{B}_{i}$ are the coeffections for the terms of the operator $H_{15}$.}}\label{tab:Pc6_weakdecay}
 \tiny
 \begin{tabular}{|lc|lc|}\hline\hline
    channel & amplitude &channel & amplitude \\\hline
    $P_{\bar csudu}^{0}\to   \pi^0   n $ & $ \frac{1}{\sqrt{2}}{(b_1-b_3-{\bar b}_{1}+{\bar b}_{3})}$&
$P_{\bar csudu}^{0}\to   \pi^-   p $ & $ -b_1-b_4+{\bar b}_{1}+{\bar b}_{4}$\\
$P_{\bar csudu}^{0}\to   K^+   \Sigma^- $ & $ -b_2+b_3+{\bar b}_{2}+{\bar b}_{3}$&
$P_{\bar csudu}^{0}\to   K^0   \Lambda^0 $ & $ \frac{2 b_1-b_2+b_4-2 {\bar b}_{1}+{\bar b}_{2}+{\bar b}_{4}}{\sqrt{6}}$\\
$P_{\bar csudu}^{0}\to   K^0   \Sigma^0 $ & $ \frac{b_2+b_4-{\bar b}_{2}+{\bar b}_{4}}{\sqrt{2}}$&
$P_{\bar csudu}^{0}\to   \eta_q   n $ & $ \frac{-b_1+2 b_2-b_3+{\bar b}_{1}-2 {\bar b}_{2}+{\bar b}_{3}}{\sqrt{6}}$\\
$P_{\bar cdsdu}^{-}\to   \pi^-   n $ & $ -b_3-b_4+{\bar b}_{3}+{\bar b}_{4}$&
$P_{\bar cdsdu}^{-}\to   K^0   \Sigma^- $ & $ b_3+b_4+{\bar b}_{3}+{\bar b}_{4}$\\
$P_{\bar cdssu}^{-}\to   \pi^0   \Sigma^- $ & $ \frac{-b_1+b_2-{\bar b}_{1}+{\bar b}_{2}}{\sqrt{2}}$&
$P_{\bar cdssu}^{-}\to   \pi^-   \Lambda^0 $ & $ \frac{b_1+b_2+2 b_4+{\bar b}_{1}+{\bar b}_{2}-2 {\bar b}_{4}}{\sqrt{6}}$\\
$P_{\bar cdssu}^{-}\to   \pi^-   \Sigma^0 $ & $ \frac{b_1-b_2+{\bar b}_{1}-{\bar b}_{2}}{\sqrt{2}}$&
$P_{\bar cdssu}^{-}\to   K^0   \Xi^- $ & $ b_1+b_4+{\bar b}_{1}+{\bar b}_{4}$\\
$P_{\bar cdssu}^{-}\to   K^-   n $ & $ b_2-b_3+{\bar b}_{2}+{\bar b}_{3}$&
$P_{\bar cdssu}^{-}\to   \eta_q   \Sigma^- $ & $ \frac{b_1+b_2-2 b_3+{\bar b}_{1}+{\bar b}_{2}-2 {\bar b}_{3}}{\sqrt{6}}$\\\hline
${P}_{\bar cdudu}^{0}\to   K^0   n $ & $ B_4\!+\!B_5\!+\!B_6\!+\!B_7\!+\!\bar{B}_4\!+\!\bar{B}_5\!+\!\bar{B}_6\!+\!\bar{B}_7$&
${P}_{\bar csudu}^{\prime 0}\to   \pi^0   n $ & $ \!\frac{-1}{\sqrt{2}}\!{(B_1\!\!+\!\!B_3\!\!+\!\!B_4\!\!+\!\!B_6\!\!+\!\!B_7\!\!-\!\!\bar{B}_1\!\!+\!\!\bar{B}_3\!\!+\!\!\bar{B}_4\!\!+\!\!\bar{B}_6\!\!-\!\!\bar{B}_7)}$\\
${P}_{\bar csudu}^{\prime 0}\to   \pi^-   p $ & $ B_1\!+\!B_3\!-\!B_5\!-\!\bar{B}_1\!+\!\bar{B}_3\!+\!\bar{B}_5$&
${P}_{\bar csudu}^{\prime 0}\to   K^+   \Sigma^- $ & $ B_1\!\!+\!\!B_2\!-\!B_6\!\!+\!\!\bar{B}_1\!\!+\!\!\bar{B}_2\!\!+\!\!\bar{B}_6$\\
${P}_{\bar csudu}^{\prime 0}\to   K^0   \Lambda^0 $ & $ \frac{1}{\sqrt{6}}{(B_4\!\!-\!\!B_1\!\!+\!\!B_2\!\!-\!\!2 \!B_3\!\!+\!\!3\!B_5\!\!+\!\!B_7\!\!-\!\!\bar{B}_1\!\!+\!\!\bar{B}_2\!\!-\!\!2\!\bar{B}_3\!\!+\!\!3\!\bar{B}_4\!\!+\!\!3\!\bar{B}_5\!\!+\!\!\bar{B}_7)}$&
${P}_{\bar csudu}^{\prime 0}\to   K^0   \Sigma^0 $ & $ \!-\!\frac{1}{\sqrt{2}}{(B_1\!\!+\!\!B_2\!\!+\!\!B_4\!\!+\!\!B_5\!\!+\!\!B_7\!\!+\!\!\bar{B}_1\!\!+\!\!\bar{B}_2\!\!-\!\!\bar{B}_4\!\!+\!\!\bar{B}_5\!\!+\!\!\bar{B}_7)}$\\
${P}_{\bar csudu}^{\prime 0}\to   \eta_q   n $ & $ \frac{1}{\sqrt{6}}{(B_4\!\!-\!\!B_1\!\!-\!\!2\!B_2\!\!+\!\!B_3\!\!+\!\!3\!B_6\!\!+\!\!B_7\!\!+\!\!\bar{B}_1\!\!-\!\!2\!\bar{B}_2\!\!+\!\!\bar{B}_3\!\!+\!\!\bar{B}_4\!\!+\!\!3\!\bar{B}_6\!\!+\!\!3\!\bar{B}_7)}$&
${P}_{\bar cdsdu}^{\prime-}\to   \pi^-   n $ & $ B_4\!+\!B_5\!+\!B_6\!+\!B_7\!+\!\bar{B}_4\!\!-\!\!\bar{B}_5\!+\!\bar{B}_6\!\!-\!\!\bar{B}_7$\\
${P}_{\bar cdsdu}^{\prime-}\to   K^0   \Sigma^- $ & $B_4\!+\!B_5\!+\!B_6\!+\!B_7\!\!-\!\!\bar{B}_4\!+\bar{B}_5\!\!-\!\bar{B}_6\!+\bar{B}_7$&
${P}_{\bar csusu}^{0}\to   \pi^+   \Sigma^- $ & $ \!-\!B_3\!-\!B_6\!+\!\bar{B}_3\!+\!\bar{B}_6$\\
${P}_{\bar csusu}^{0}\to   \pi^0   \Lambda^0 $ & $ \!-\!\frac{2 B_1\!+\!B_2\!+\!B_3\!+\!B_4\!+\!B_7\!-\!2 \bar{B}_1\!-\!\bar{B}_2\!-\!\bar{B}_3\!+\!3 \bar{B}_4\!-\!\bar{B}_7}{2 \sqrt{3}}$&
${P}_{\bar csusu}^{0}\to   \pi^0   \Sigma^0 $ & $ \frac{\!-\!B_2\!-\!B_3\!+\!B_4\!+\!B_7\!+\!\bar{B}_2\!+\!\bar{B}_3\!-\!\bar{B}_4\!-\!\bar{B}_7}{2}$\\
${P}_{\bar csusu}^{0}\to   \pi^-   \Sigma^+ $ & $ \!-\!B_2\!-\!B_5\!+\!\bar{B}_2\!+\!\bar{B}_5$&
${P}_{\bar csusu}^{0}\to   K^+   \Xi^- $ & $ \!-\!B_1\!-\!B_2\!-\!B_3\!-\!\bar{B}_1\!-\!\bar{B}_2\!+\!\bar{B}_3$\\
${P}_{\bar csusu}^{0}\to   K^0   \Xi^0 $ & $ \!-\!B_2\!-\!B_5\!-\!\bar{B}_2\!-\!\bar{B}_5$&
${P}_{\bar csusu}^{0}\to   \overline K^0   n $ & $ \!-\!B_3\!-\!B_6\!-\!\bar{B}_3\!-\!\bar{B}_6$\\
${P}_{\bar csusu}^{0}\to   K^-   p $ & $ \!-\!B_1\!-\!B_2\!-\!B_3\!+\!\bar{B}_1\!+\!\bar{B}_2\!-\!\bar{B}_3$&
${P}_{\bar csusu}^{0}\to   \eta_q   \Lambda^0 $ & $ \frac{B_4\!-\!4 B_1\!-\!5 B_2\!-\!5 B_3\!+\!B_7\!-\!3 \bar{B}_2\!-\!3 \bar{B}_3\!+\!3 \bar{B}_4\!+\!3 \bar{B}_7}{6}$\\
${P}_{\bar csusu}^{0}\to   \eta_q   \Sigma^0 $ & $ \!-\!\frac{2 B_1\!+\!B_2\!+\!B_3\!+\!B_4\!+\!B_7\!+\!2 \bar{B}_1\!-\!\bar{B}_2\!-\!\bar{B}_3\!-\!\bar{B}_4\!+\!3 \bar{B}_7}{2 \sqrt{3}}$&
${P}_{\bar cdssu}^{\prime-}\to   \pi^0   \Sigma^- $ & $ \frac{B_2\!-\!B_3\!-\!B_4\!-\!2 B_6\!-\!B_7\!-\!\bar{B}_2\!+\!\bar{B}_3\!+\!\bar{B}_4\!+\!2 \bar{B}_6\!+\!\bar{B}_7}{\sqrt{2}}$\\
${P}_{\bar cdssu}^{\prime-}\to   \pi^-   \Lambda^0 $ & $ \frac{2 B_1\!+\!B_2\!+\!B_3\!+\!B_4\!+\!B_7\!-\!2 \bar{B}_1\!-\!\bar{B}_2\!-\!\bar{B}_3\!+\!3 \bar{B}_4\!-\!\bar{B}_7}{\sqrt{6}}$&
${P}_{\bar cdssu}^{\prime-}\to   \pi^-   \Sigma^0 $ & $ \frac{B_3\!-\!B_2\!-\!B_4\!-\!2 B_5\!-\!B_7\!+\!\bar{B}_2\!-\!\bar{B}_3\!+\!\bar{B}_4\!+\!2 \bar{B}_5\!+\!\bar{B}_7}{\sqrt{2}}$\\
${P}_{\bar cdssu}^{\prime-}\to   K^0   \Xi^- $ & $ B_1\!+\!B_3\!-\!B_5\!+\!\bar{B}_1\!-\!\bar{B}_3\!-\!\bar{B}_5$&
${P}_{\bar cdssu}^{\prime-}\to   K^-   n $ & $ B_1\!+\!B_2\!-\!B_6\!-\!\bar{B}_1\!-\!\bar{B}_2\!-\!\bar{B}_6$\\
${P}_{\bar cdssu}^{\prime-}\to   \eta_q   \Sigma^- $ & $ \frac{1}{\sqrt{6}}{(2 B_1\!\!+\!B_2\!\!+\!\!B_3\!\!+\!B_4\!\!+\!\!B_7\!\!+\!\!2 \bar{B}_1\!\!-\!\!\bar{B}_2\!\!-\!\!\bar{B}_3\!\!-\!\!\bar{B}_4\!+\!3 \bar{B}_7)}$&
${P}_{\bar cdsds}^{--}\to   \pi^-   \Sigma^- $ & $ B_4\!+\!B_5\!+\!B_6\!+\!B_7\!-\!\bar{B}_4\!-\!\bar{B}_5\!-\!\bar{B}_6\!-\!\bar{B}_7$\\\hline
\hline
\end{tabular}
\end{table}
\section{Productions from $b$ hardrons}\label{sec:production}
In this section we investigate the production of the pentaquark states from the $b$ hardrons. When the production is calculated with SU(3) symmetry analysis, the representation of the initial and final states is an essential input.
The initial states of $b$ hardrons include $B$ meson and $b$ baryon.
The representations of the anti-light baryons can be obtained from those of the light baryon~\cite{Xing:2018bqt,Wang:2017azm}. The current $b$ factories, such as the LHC in Europe and SuperKEKB in Japan, have achieved significant progress in studying exotic hadrons. The Belle detector at KEKB and the BABAR detector at PEP-II have integrated luminosities of about $988~\rm{fb}^{-1}$ and $518~\rm{fb}^{-1}$, respectively~\cite{Li:2018dhh,Belle-II:2018jsg}. These detectors mainly operate at the energy of the $\varUpsilon (4S)$ resonance. In the future, Belle II is expected to record $50~\rm{ab}^{-1}$ of data~\cite{Belle-II:2022cgf}. Since $\varUpsilon (4S)$ can decay into a $B\bar{B}$ pair, this will provide ample opportunities to study $B$ meson and $b$ baryon.

\subsection{Productions from $B$ mesons}
\begin{figure}
\includegraphics[width=0.85\columnwidth,height=0.35\textheight]{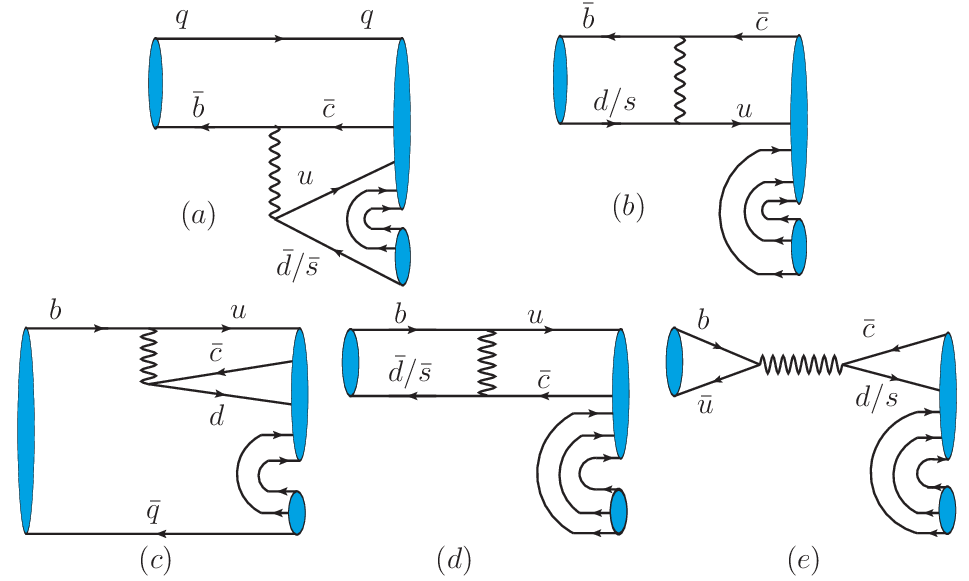}
\caption{The production topographies of the singly anti-charmed pentaquark ground state, starting from the $B$ mesons $b\bar{q}$ and $\bar{b}q$. Diagrams (a) and (b) show the production processes of the singly anti-charmed pentaquark multiple states and the anti-light baryons from the $B$ meson decays $\bar{b}q$, all of which are Cabibbo allowed. Diagrams (c-e) show those from the anti-B meson $b\bar{q}$, all of which are Cabibbo-suppressed.  For them the different decay mechanisms are involved: for the transition $\bar{b}\to \bar{c}u\bar{d}/\bar{s}$ the rates of the different decay widths are given as follows: outer W emission (a), W exchange (b, d), inner W emission (c) and annihilation (e).}
\label{fig:production}
\end{figure}

The production of the ground states can be achieved by $B$ meson weak decays, which can be induced by $\bar{b}\to \bar{c}u\bar{d}/\bar{s}$ and $b\to u \bar{c}d/s$. We write down the possible Hamiltonian for the production of the pentaquark ground states directly from a $B$ meson {$(\bar{b}q)$ noted with $B^{i}=\epsilon^{ijk}B_{[jk]}$ or an anti-$B$ meson ($b\bar{q}$) denoted by $\overline{B}_{i}$},
{\begin{eqnarray}
 {\mathcal{H}}_{\text{p}}
   & =&c_{1}{B}^{i}(H_{\boldsymbol{8}})_{i}^{j}(\overline{P_{\bar{c}\boldsymbol{3}}})_{k}(P_{\bar{8}})_{j}^{k}
  + c_{2}{B}^{i}(H_{\boldsymbol{8}})_{k}^{j}(\overline{P_{\bar{c}\boldsymbol{3}}})_{i}(P_{\bar{8}})_{j}^{k} 
  + c_{3}{B}^{i}(H_{\boldsymbol{8}})_{k}^{j}(\overline{P_{\bar{c}\boldsymbol{3}}})_{j}(P_{\bar{8}})_{i}^{k} \nonumber\\
 &&+  
  \bar{c}_{1}\overline{B}_{i}(H_{\boldsymbol{\bar{3}}})^{[ij]}(\overline{P_{\bar{c}\boldsymbol{3}}})_{k}(P_{\bar{8}})_{k}^{j} 
  +  \bar{c}_{2}\overline{B}_{k}(H_{\boldsymbol{\bar{3}}})^{[ij]}(\overline{P_{\bar{c}\boldsymbol{3}}})_{i}(P_{\bar{8}})_{k}^{j}
  +\bar{c}_{3}\overline{B}_{i}(H_{\boldsymbol{6}})^{[ij]}(\overline{P_{\bar{c}\boldsymbol{3}}})_{k}(P_{\bar{8}})_{k}^{j}\nonumber\\ 
  &&
   +\bar{c}_{4}\overline{B}_{k}(H_{\boldsymbol{6}})^{[ij]}(\overline{P_{\bar{c}\boldsymbol{3}}})_{i}(P_{\bar{8}})_{k}^{j}
 + e_{1}{B}_{[ij]}(H_{\boldsymbol{8}})_{k}^{i}(\overline{{P}_{\bar{c}\boldsymbol{\bar 6}}})^{\{jl\}}(P_{\bar{8}})_{l}^{k} 
  +e_{2}{B}_{[ij]}(H_{\boldsymbol{8}})_{k}^{i}(\overline{{P}_{\bar{c}\boldsymbol{\bar 6}}})^{\{kl\}}(P_{\bar{8}})_{l}^{j} \nonumber\\
  &&
  +e_{3}{B}_{[ij]}(H_{\boldsymbol{8}})_{k}^{l}(\overline{{P}_{\bar{c}\boldsymbol{\bar 6}}})^{\{ik\}}(P_{\bar{8}})_{l}^{j} 
 + \bar{e}_{1}\overline{B}_{i}(H_{\boldsymbol{\bar{3}}})_{j}(\overline{{P}_{\bar{c}\boldsymbol{\bar 6}}})^{\{ik\}}(P_{\bar{8}})_{k}^{j} + \bar{e}_{2}\overline{B}_{i}(H_{\boldsymbol{\bar{3}}})_{j}(\overline{{P}_{\bar{c}\boldsymbol{\bar 6}}})^{\{jk\}}(P_{\bar{8}})_{k}^{i}\nonumber\\
 &&+ \bar{e}_{3}\overline{B}_{i}(H_{\boldsymbol{6}})^{\{ij\}}(\overline{{P}_{\bar{c}\boldsymbol{\bar 6}}})^{\{\alpha k\}}(P_{\bar8})_{k}^{l}\ \varepsilon_{\alpha jl}.\label{eq:hamiltonian6}
 \end{eqnarray}
 The parameters $c_{i}$, $\bar{c}_{i}$, $e_{i}$, $\bar{e}_{i}$, with $i=1,2,3,4$}, are the non-perturbative coefficients. The operator $H_{\boldsymbol{8}}$ means the transition $\bar{b}\to \bar{c}u\bar{d}/\bar{s}$ which is Cabibbo allowed, while $H_{\boldsymbol{\bar{3}}}$ and $H_{\boldsymbol{6}}$ indicate $b\to u \bar{c}d/s$~\cite{He:2016xvd} which is Cabibbo suppressed. Since the pentaquark ground states are the final states, they need to be noted with {$\overline{P_{\bar{c}\boldsymbol{3}}}$ and $\overline{P_{\bar{c}\boldsymbol{\bar 6}}}$}. At the quark level, the productions of singly charmed pentaquark multiple states can be described with Feynman diagrams, as shown in Fig.~\ref{fig:production}.
In particular, the Hamiltonian with the coefficient $e_{i}(i=1,2,3)$ corresponds to the diagrams Fig.~\ref{fig:production}.(a) and (b), by comparison, the diagrams Fig.~\ref{fig:production}.(c-e) correspond to the Hamiltonian with the coefficient {$\bar{e}_{i}(i=1,2,3)$.}

We expand the Hamiltonian Eq.~(\ref{eq:hamiltonian6}) and harvest the possible amplitude results, aggregating them in Tab.~\ref{tab:6statesp}. {Based on the amplitudes relationship, while disregarding the impact of phase-space integrals, and utilizing the mass spectrum of the pentaquark states from the Chromomagnetic Interaction model~\cite{An:2020vku} and that of the $B$ mesons and light baryons from the Particle Data Group~\cite{ParticleDataGroup:2022pth}, we can calculate the rates of various channel widths.}
For the transition $\bar{b}\to \bar{c}u\bar{d}/\bar{s}$, the rates of the different decay widths are given as follows:
{
\begin{eqnarray}
&&\Gamma_{B^0_s\to   {P}_{\bar{c}sudu}^{(\prime) 0}  \overline \Sigma^0}:\Gamma_{B^0_s\to   {P}_{\bar{c}dsdu}^{(\prime)-}  \overline\Sigma^+}:\Gamma_{B^0\to   {P}_{\bar{c}dssu}^{(\prime)-}  \overline\Xi^+}=1:2:31,\ 
  \Gamma_{B^0\to   {P}_{\bar{c}sudu}^{(\prime) 0}  \overline \Xi^0}:\Gamma_{B^0_s\to   {P}_{\bar{c}sudu}^{(\prime) 0}  \overline n}=1:21,\nonumber\\
      &&\Gamma_{ B^0\to   {P}_{\bar{c}dudu}^{0}  \overline \Sigma^0 } \!: \!\Gamma_{ B^+\to   {P}_{\bar{c}dudu}^{0}  \overline\Sigma^+ } \!: \!\Gamma_{ B^+\to   {P}_{\bar{c}sudu}^{\prime 0}  \overline\Sigma^+ }\!: \!\Gamma_{ B^+\to   {P}_{\bar{c}sudu}^{\prime 0}  \overline\Xi^+ } \!: \! \Gamma_{ B^+\to   {P}_{\bar{c}susu}^{0}  \overline\Xi^+ }= 1 \!: \!2 \!: \!36 \!: \!2 \!: \!30, \ 
 \nonumber\\
 &&  \Gamma_{ B^+\to   P_{\bar{c}sudu}^{0}  \overline\Sigma^+ }: \Gamma_{ B^+\to   P_{\bar{c}sudu}^{0}  \overline\Xi^+ } = 18:1,\ 
  \Gamma_{B^0_s\to   P_{\bar{c}dssu}^{-}  \overline\Xi^+}:\Gamma_{B^0\to   P_{\bar{c}dsdu}^{-}  \overline\Sigma^+}=1:22,\nonumber\\
    && \Gamma_{B^0\to   P_{\bar{c}dsdu}^{-}  \overline\Xi^+}:\Gamma_{B^0_s\to   P_{\bar{c}dssu}^{-}  \overline\Sigma^+}=1:18, \Gamma_{B^0_s\to   {P}_{\bar{c}dudu}^{0}  \overline n}:\Gamma_{B^0\to   {P}_{\bar{c}susu}^{0}  \overline \Xi^0}=14:1,\nonumber\\
      &&\Gamma_{B^0_s\to   {P}_{\bar{c}susu}^{0}  \overline \Xi^0}:\Gamma_{B^0_s\to   {P}_{\bar{c}dssu}^{\prime-}  \overline\Xi^+}:\Gamma_{B^0\to   {P}_{\bar{c}dudu}^{0}  \overline n} :\Gamma_{B^0\to   {P}_{\bar{c}dsdu}^{\prime-}  \overline\Sigma^+}=1:1:25:22,\nonumber\\
    && \Gamma_{B^0\to   {P}_{\bar{c}dsdu}^{\prime-}  \overline\Xi^+}:\Gamma_{B^0_s\to   {P}_{\bar{c}susu}^{0}  \overline \Sigma^0}:\Gamma_{B^0_s\to   {P}_{\bar{c}dssu}^{\prime-}  \overline\Sigma^+}=1:9:18.
\end{eqnarray}}

This shows that the difference between the decay widths of different production processes is relatively large but related. As soon as any decay channel is detected in the future, we can give other decay widths. And if two decay widths can be measured, our prediction can also be verified. Typically, the Cabibbo allowed (CA) production channels, which are mainly induced by the transition $\bar{b}\to \bar{c}u\bar{d}/\bar{s}$, can receive the largest contribution. In addition, the Cabibbo suppressed (CS) channels $b\to u \bar{c}d/s$ are shown with the diagrams (c-e) in Fig.~\ref{fig:production}. Furthermore, since the detection efficiencies of the light baryons are different, the production channel containing an efficiently detected baryon may have a larger number of production events of the pentaquark states. In this article we prefer the channels with $N$, $\Lambda(\to p\pi^-)$, $\Sigma^0(\to \Lambda\gamma)$, $\Sigma^+(\to p\pi^0)$ and $\Xi^+(\to \Lambda\pi^+)$ to those with other light baryons. Consequently, we extract the golden channel for each ground state ({${P}_{\bar{c}dudu}^{0}$,~$P_{\bar{c}sudu}^{(\prime) 0}$, ${P}_{\bar{c}dsdu}^{(\prime)-}$, ${P}_{\bar{c}dssu}^{(\prime)-}$, ${P}_{\bar{c}susu}^{0}$ and $P_{\bar{c}dsds}^{--}$}), given by Eq.~(\ref{eq:goldenBdecay}), and propose them to study in the future experiment.
{
\begin{eqnarray}
  CA:&&B^0_s \to   {P}_{\bar{c}susu}^{0}  \overline \Lambda, 
  B^0_s \to   {P}_{\bar{c}susu}^{0}  \overline \Sigma^0, 
  B^+ \to    {P}_{\bar{c}susu}^{0}  \overline\Xi^+, 
  B^0 \to   {P}_{\bar{c}dudu}^{0}  \overline n,
  B^0_s \to   {P}_{\bar{c}sudu}^{(\prime) 0}  \overline n, 
 \nonumber\\
  &&
  B^+\to   {P}_{\bar{c}sudu}^{(\prime) 0}   \overline \Sigma^+,    
  B^0 \to   {P}_{\bar{c}sudu}^{(\prime) 0}  \overline \Lambda, 
  B^0 \to   {P}_{\bar{c}sudu}^{(\prime) 0}  \overline \Sigma^0, \nonumber\\
  &&
  B^0 \to   {P}_{\bar{c}dsdu}^{(\prime)-}  \overline\Sigma^+, 
  B^0\!\! \to\!\!   {P}_{\bar{c}dssu}^{(\prime)-}  \overline\Xi^+, 
  B^0_s\to   {P}_{\bar{c}dssu}^{(\prime)-}  \overline \Sigma^+.\label{eq:goldenBdecay}
\end{eqnarray}}

Among them, the state {$P_{\bar{c}dsds}^{--}$} cannot be produced by the Cabibbo-allowed channels, since the states generated by these channels contain at least one $u$ quark, as shown with diagrams (a) and (b) in Fig.~\ref{fig:production}.
In the next section, we will discuss the production modes for these ground states. We list the relationship between decay widths, ignoring the phase space differences, in the following Eq.~(\ref{eq:relation6bp}).
{
\begin{eqnarray}
  CA:&&\Gamma( B^0_s\to P_{\bar csudu}^{0} \overline \Sigma^0)= \frac{1}{2}\Gamma( B^0_s\to P_{\bar cdsdu}^{-} \overline \Sigma^+), \Gamma( B^+\!\to\! {P}_{\bar{c}sudu}^{\prime 0} \overline\Sigma^+)= \Gamma( B^+\!\to\! {P}_{\bar{c}susu}^{0} \overline\Xi^+),\nonumber\\
    &&\Gamma( B^0\!\to\! {P}_{\bar{c}dudu}^{0} \overline n)= \Gamma( B^0\!\to\! {P}_{\bar{c}dsdu}^{\prime-} \overline\Sigma^+),\ 
  \Gamma( B^0_s\!\to\! {P}_{\bar{c}sudu}^{\prime 0} \overline \Sigma^0)= \frac{1}{2}\Gamma( B^0_s\!\to\! {P}_{\bar{c}dsdu}^{\prime-} \overline\Sigma^+),\nonumber\\
  &&\Gamma( B^0_s\!\to\! {P}_{\bar{c}susu}^{0} \overline \Sigma^0)= \frac{1}{2}\Gamma( B^0_s\!\to\! {P}_{\bar{c}dssu}^{\prime-} \overline\Sigma^+),\ 
  \Gamma( B^0_s\!\to\! {P}_{\bar{c}susu}^{0} \overline \Xi^0)= \Gamma( B^0_s\!\to\! {P}_{\bar{c}dssu}^{\prime-} \overline\Xi^+),
  \nonumber\\
  &&\Gamma( B^+\!\to\! {P}_{\bar{c}dudu}^{0} \overline\Sigma^+)= \Gamma( B^+\!\to\! {P}_{\bar{c}sudu}^{\prime 0} \overline\Xi^+)= 2\Gamma( B^0\!\to\! {P}_{\bar{c}dudu}^{0} \overline \Sigma^0).\label{eq:relation6bp}
    \end{eqnarray}}

    Similarly, for the exotic states $\boldsymbol{15}$ and $\boldsymbol{15^{\prime}}$, their production channels and relationships of decay widths, can also be obtained by expanding the corresponding possible Hamiltonian. For convenience, the rest of Hamiltonian and the decay width relations between different production channels are also reorganized in Appendix~\ref{sec:appendix}. The possible Hamiltonian has been given with Eq.~(\ref{eq:hamiltonian615p}).

\begin{table}
\caption{{The production from B mesons of the pentaquark ground states.}}\label{tab:6statesp}\begin{tabular}{|lc|lc|lc|c|c}\hline\hline
channel & amplitude &channel & amplitude&channel & amplitude\\\hline
$ B^+\to   P_{\bar csudu}^{0}  \overline \Sigma^+ $ & $ (c_2+c_3) V_{{ud}}$&
$ B^+\to   P_{\bar csudu}^{0}  \overline \Xi^+ $ & $ (c_2+c_3) V_{{us}}$&
$ B^0\to   P_{\bar csudu}^{0}  \overline \Lambda^0 $ & $ \frac{(c_1+c_3) V_{{ud}}}{\sqrt{6}}$\\
$ B^0\to   P_{\bar csudu}^{0}  \overline \Sigma^0 $ & $ \frac{(c_1-c_3) V_{{ud}}}{\sqrt{2}}$&
$ B^0\to   P_{\bar csudu}^{0}  \overline \Xi^0 $ & $ c_3 V_{{us}}$&
$ B^0\to   P_{\bar cdsdu}^{-}  \overline \Sigma^+ $ & $ (c_1+c_2) V_{{ud}}$\\
$ B^0\to   P_{\bar cdsdu}^{-}  \overline \Xi^+ $ & $ c_2 V_{{us}}$&
$ B^0\to   P_{\bar cdssu}^{-}  \overline \Xi^+ $ & $ c_1 V_{{ud}}$&
$ B^0_s\to   P_{\bar csudu}^{0}  \overline \Lambda^0 $ & $ \frac{(c_1-2 c_3) V_{{us}}}{\sqrt{6}}$\\
$ B^0_s\to   P_{\bar csudu}^{0}  \overline \Sigma^0 $ & $ \frac{c_1 V_{{us}}}{\sqrt{2}}$&
$ B^0_s\to   P_{\bar csudu}^{0}  \overline n $ & $ c_3 V_{{ud}}$&
$ B^0_s\to   P_{\bar cdsdu}^{-}  \overline \Sigma^+ $ & $ c_1 V_{{us}}$\\
$ B^0_s\to   P_{\bar cdssu}^{-}  \overline \Sigma^+ $ & $ c_2 V_{{ud}}$&
$ B^0_s\to   P_{\bar cdssu}^{-}  \overline \Xi^+ $ & $ (c_1+c_2) V_{{us}}$&&\\\hline
  $ B^+\!\! \to\!\!   {P}_{\bar{c}dudu}^{0}  \overline\Sigma^+ $ & $ -e_3 V_{{us}}$&
  $ B^+\!\! \to\!\!    {P}_{\bar{c}sudu}^{\prime 0}  \overline\Sigma^+ $ & $ -e_3 V_{{ud}}$&
  $ B^+\!\! \to\!\!    {P}_{\bar{c}sudu}^{\prime 0}  \overline\Xi^+ $ & $ e_3 V_{{us}}$\\
  $ B^+\!\! \to \!\!   {P}_{\bar{c}susu}^{0}  \overline\Xi^+ $ & $ e_3 V_{{ud}}$&
  $ B^0\!\! \to\!\!    {P}_{\bar{c}dudu}^{0}  \overline \Lambda $ & $ \frac{(2 e_1\!+\!2 e_2\!+\!e_3) V_{{us}}}{\sqrt{6}}$&
  $ B^0\!\! \to\!\!    {P}_{\bar{c}dudu}^{0}  \overline \Sigma^0 $ & $ \frac{e_3 V_{{us}}}{\sqrt{2}}$\\
  $ B^0\!\! \to\!\!   {P}_{\bar{c}dudu}^{0}  \overline n $ & $ -e_1 V_{{ud}}$&
  $ B^0\!\! \to\!\!   {P}_{\bar{c}sudu}^{\prime 0}  \overline \Lambda $ & $ \frac{(2 e_2- e_1+e_3) V_{{ud}}}{\sqrt{6}}$&
  $ B^0\!\! \to\!\!   {P}_{\bar{c}sudu}^{\prime 0}  \overline \Sigma^0 $ & $ \frac{(e_1+e_3) V_{{ud}}}{\sqrt{2}}$\\
  $ B^0\!\! \to\!\!   {P}_{\bar{c}sudu}^{\prime 0}  \overline \Xi^0 $ & $ -(e_1\!+\!e_2) V_{{us}}$&$ B^0\!\! \to\!\!   {P}_{\bar{c}dsdu}^{\prime-}  \overline\Xi^+ $ & $ -(e_1\!\!+\!\!e_2\!\!+\!\!e_3) V_{{us}}$&
  $ B^0\!\! \to\!\!   {P}_{\bar{c}dsdu}^{\prime-}  \overline\Sigma^+ $ & $ -e_1 V_{{ud}}$
  \\
  $ B^0\!\! \to\!\!   {P}_{\bar{c}susu}^{0}  \overline \Xi^0 $ & $ -e_2 V_{{ud}}$&
  $ B^0\!\! \to\!\!   {P}_{\bar{c}dssu}^{\prime-}  \overline\Xi^+ $ & $ -(e_2\!\!+\!\!e_3) V_{{ud}}$&
  $ B^0_s\!\! \to\!\!   {P}_{\bar{c}dudu}^{0}  \overline n $ & $ e_2 V_{{us}}$\\
  $ B^0_s\!\! \to\!\!   {P}_{\bar{c}sudu}^{\prime 0}  \overline \Lambda $ & $ \frac{(e_2-2 e_1-e_3) V_{{us}}}{\sqrt{6}}$&
  $ B^0_s\!\! \to\!\!   {P}_{\bar{c}sudu}^{\prime 0}  \overline \Sigma^0 $ & $ -\frac{(e_2+e_3) V_{{us}}}{\sqrt{2}}$&
  $ B^0_s\!\! \to\!\!   {P}_{\bar{c}sudu}^{\prime 0}  \overline n $ & $ (e_1\!\!+\!\!e_2) V_{{ud}}$\\
  $ B^0_s\!\! \to\!\!   {P}_{\bar{c}dsdu}^{\prime-}  \overline\Sigma^+ $ & $ (e_2+e_3) V_{{us}}$&
  $ B^0_s\!\! \to\!\!   {P}_{\bar{c}susu}^{0}  \overline \Lambda $ & $ \frac{(e_1\!+\!e_2\!-\!e_3) V_{{ud}}}{\sqrt{6}}$&$ B^0_s\!\! \to\!\!   {P}_{\bar{c}susu}^{0}  \overline \Xi^0 $ & $ e_1 V_{{us}}$
  \\
  $ B^0_s\!\! \to\!\!   {P}_{\bar{c}susu}^{0}  \overline \Sigma^0 $ & $ -\frac{(e_1+e_2+e_3) V_{{ud}}}{\sqrt{2}}$&
  $ B^0_s\!\! \to\!\!   {P}_{\bar{c}dssu}^{\prime-}  \overline\Sigma^+ $ & $ (e_1\!\!+\!\!e_2\!\!+\!\!e_3) V_{{ud}}$&
  $ B^0_s\!\! \to\!\!   {P}_{\bar{c}dssu}^{\prime-}  \overline\Xi^+ $ & $ e_1 V_{{us}}$\\\hline\hline
  \end{tabular}
  \end{table}

  \subsection{Productions from $b$ baryon}

  In our previous work~\cite{Xing:2022aij}, we studied the production of the singly anti-charmed pentaquark ground states {$P_{\bar{c}\boldsymbol{3}}$ and $P_{\bar{c}\boldsymbol{\bar 6}}$} from bottomed baryon decay. For each ground state, several golden decay modes can be selected from Cabibbo allowed channels, which can be easily reconstructed from their corresponding final states. At the quark level, these decays involve the dominant weak process $b\to u \bar{c}d/s$.
 {
  \begin{eqnarray}\label{eq:Cabibbogoldenbbaryon}
  &&\Lambda_{b}^{0}\to {P}_{\bar{c}dudu}^{0}\bar{K}^{0},~\Lambda_{b}^{0}\to {P}_{\bar{c}sudu}^{(\prime) 0}{K}^{-},~\Lambda_{b}^{0}\to {P}_{\bar{c}dsdu}^{(\prime)-}\pi^{+},\nonumber\\
  &&~\Xi_{b}^{0}\to {P}_{\bar{c}dssu}^{(\prime)-}\pi^{+},~\Xi_{b}^{-}\to P_{\bar{c}dsds}^{--}\pi^{+},~\Xi_{b}^{-}\to {P}_{\bar{c}susu}^{0}\pi^{-}.
  \end{eqnarray}}
  \begin{table}
    \caption{{Pentaquark signatures in bottomed baryon decays and reconstruction through the strong and weak decays. The $S.1$ and $S.2$ represent the reconstruction through strong and weak decays respectively.}}
    \label{tab:baryonpro}
    \begin{tabular}{c|c|c|c}
      \hline\hline
      $\qquad\qquad$   &ground states&production mode&experimental signatures \\\hline 
     \multirow{5}{*}{S.1    } &${P}_{\bar{c}dudu}^{0}$&$\Lambda_{b}^{0}\to {P}_{\bar{c}dudu}^{0}(\to D^{-}p)\bar{K}^{0}$&M($D^{-}p$)\\
      &$P_{\bar{c}sudu}^{(\prime) 0}$&$\Lambda_{b}^{0}\to {P}_{\bar{c}sudu}^{(\prime) 0}(\to D^{-}_{s}p){K}^{-}$&M($D^{-}_{s}p$)\\
      &${P}_{\bar{c}dsdu}^{(\prime)-}$&$\Lambda_{b}^{0}\to {P}_{\bar{c}dsdu}^{(\prime)-}(\to D^{-}\Lambda)\pi^{+}$&M($D^{-}\Lambda$)\\
      &${P}_{\bar{c}dssu}^{(\prime)-}$&$\Xi_{b}^{0}\to {P}_{\bar{c}dssu}^{(\prime)-}(\to D^{-}_{s}\Sigma^{0})\pi^{+}$&M($D^{-}_{s}\Sigma^{0}$)\\
      &$P_{\bar{c}dsds}^{--}$&$\Xi_{b}^{-}\to P_{\bar{c}dsds}^{--}(\to D^{-}\Xi^{-})\pi^{+}$&M($D^{-}\Xi^{-}$)\\\hline
    \multirow{3}{*}{S.2}&$P_{\bar{c}sudu}^{(\prime) 0}$&$\Lambda_{b}^{0}\to {P}_{\bar{c}sudu}^{(\prime) 0}(\to \pi^{-} p){K}^{-}$& M$(\pi^{-} p)$\\
    &${P}_{\bar{c}dssu}^{(\prime)-}$&$\Xi_{b}^{0}\to {P}_{\bar{c}dssu}^{(\prime)-}(\to \pi^{-}\Lambda^{0})\pi^{+}$&M$(\pi^{-}\Lambda^{0})$\\
    &${P}_{\bar{c}susu}^{0}$ & $\Xi_{b}^{-}\to {P}_{\bar{c}susu}^{0}(\to K^{-} p)\pi^{-}$& M$(K^{-} p)$\\\hline\hline
    \end{tabular}
  \end{table}

The BelleII collider is expected to produce $5.4\times10^{10}$ $B\overline B^{0}/B^{+}B^{-}$ pairs and $6\times10^{8}$ $B_{s}^{0}\overline B_{s}^{0}$ pairs when the integrated luminosity at $\varUpsilon(4S)$ and $\varUpsilon(5S)$ is $50~ab^{-1}$ and $5~ab^{-1}$, respectively~\cite{Wang:2022nrm,Belle-II:2018jsg,HFLAV:2019otj}. 
Moreover, KEKB and PEP-II can provide a large number of charm samples, including those produced from $B$ meson decays via $b\to c$ and many collision-produced samples with good reconstruction efficiency, offering more opportunities to study the anti-charmed pentaquark. Belle II is uniquely positioned to investigate heavy exotic hadrons~\cite{Belle-II:2018jsg,Belle-II:2022cgf}.
Fruitful results on charm physics have been achieved by the LHCb experiments, based on data sets with an integrated luminosity of about $9.6fb^{-1}$~\cite{Li:2018dhh,LHCb:2017vec,LHCb:2022syj}. In the future, the integrated luminosity of LHCb will reach $300fb^{-1}$, there will be $3\times 10^{13}$ pairs of $B\overline B^{0}/B^{+}B^{-}$, $1\times 10^{13}$ pairs of
$B_{s}^{0}\overline B_{s}^{0}$, and $2\times 10^{13}$ pairs of $\Lambda_{b}\overline{\Lambda}_{b}$~\cite{Wang:2022nrm,LHCb:2018roe,LHCb:2016qpe,LHCb:2019fns,HFLAV:2019otj,LHCb:2019tea}. 

By estimating the magnitudes of the CKM matrix elements, it is possible to determine that the production of the pentaquark ground state from bottom quark decay is dominant.
The typical branching fraction of bottom quark decay may be less than $10^{-3}$. In addition, the reconstruction of the final charm pentaquark state requires another factor of $10^{-3}$ and a few percent for the reconstruction of the light baryon. Therefore, the branching ratio of the exotic state generated by the $B\bar{B}$ pairs is expected to be large enough, on the order of $10^{-8}$, to give a good chance of finding it. We calculate the production of the pentaquark ground state from the weak decay of the $b$ hadrons, taking into account the weak and strong decays of them.
To find them experimentally, some golden channels have been selected and arranged by Eq.~(\ref{eq:golden}), involving the dominant weak process $\bar{b}\to \bar{c}u\bar{d}/\bar{s}$ and $b\to u \bar{c}d/s$.
After considering the corresponding weak and strong decays of the pentaquark ground states, the golden modes for the reconstruction of them are collected in Tab.~\ref{tab:swdecay1}.

\begin{table}
  \caption{{Pentaquark signatures in $B$ meson decays and reconstruction through the strong and weak decays. The $S.1$ and $S.2$ represent the reconstruction through strong and weak decays respectively.}}\label{tab:swdecay1}
  \begin{tabular}{c|c|c|c}
    \hline\hline
    $\qquad\qquad$ &ground states&production mode&\makecell[c]{experimental \\signatures}\\\hline 
    \multirow{4}{*}{S.1}&{${P}_{\bar{c}dudu}^{0}$}&${B}^{0}\!\!\to\!\! {P}_{\bar{c}dudu}^{0}(\!\!\to\!\! D^{-}p)\bar{n}$&M($D^{-}p$)\\
   &{${P}_{\bar{c}dssu}^{(\prime)-}$}&${B}^{0}\!\!\to\!\! {P}_{\bar{c}dssu}^{(\prime)-}(\!\!\to\!\! D^{-}_{s}\Sigma^{0})\overline{\Xi}^{+}$&M($D^{-}_{s}\Sigma^{0}$)\\
   &{$P_{\bar{c}sudu}^{(\prime) 0}$}&${B}^{0}\!\!\to\!\! {P}_{\bar{c}sudu}^{(\prime) 0}(\!\!\to\!\! D^{-}_{s}p)\overline{\Sigma}^{0}$&M($D^{-}_{s}p$)\\
  &{${P}_{\bar{c}dsdu}^{(\prime)-}$}&${B}^{0}\!\!\to\!\! {P}_{\bar{c}dsdu}^{(\prime)-}(\!\!\to\!\! D^{-}\Lambda)\overline{\Sigma}^{+}$&M($D^{-}\Lambda$)\\ 
  \hline
  \multirow{3}{*}{S.2}&{${P}_{\bar{c}susu}^{0}$}&${B}^{+}\!\!\to\!\! {P}_{\bar{c}susu}^{0}(\!\!\to\!\! K^{-} p)\overline{\Xi}^{+}$& M$(K^{-} p)$\\
  &{$P_{\bar{c}sudu}^{(\prime) 0}$}&${B}^{0}\!\!\to\!\! {P}_{\bar{c}sudu}^{(\prime) 0}(\!\!\to\!\! \pi^{-} p)\overline{\Sigma}^{0}$& M$(\pi^{-} p)$\\
  &{${P}_{\bar{c}dssu}^{(\prime)-}$}&${B}^{0}\!\!\to\!\! {P}_{\bar{c}dssu}^{(\prime)-}(\!\!\to\!\! \pi^{-}\Lambda^{0})\overline{\Sigma}^{+}$&M$(\pi^{-}\Lambda^{0})$\\ \hline\hline
  \end{tabular}
\end{table}

\section{Summary}\label{sec:summary}
In this discussion, we will explore the production of the ground state of the singly anti-charmed pentaquark through B decays using the light quark SU(3) symmetry analysis. To construct the possible Hamiltonian of production for the ground states, we will use the representations of $B$ mesons and final states ({$P_{\bar{c}\boldsymbol{3}}$, $P_{\bar{c}\boldsymbol{\bar 6}}$} and $P_{8}$) given by Eq.~(\ref{eq:hamiltonian6}). We will obtain the amplitude of different channels, as given by Tab.~\ref{tab:6statesp}, and also collect the relations between different channels as mentioned in Eq.~(\ref{eq:relation6bp}). To ensure clarity, we will suggest several dominant channels for the search for the singly anti-charmed pentaquark ground states in b-factory experiments. 
{
\begin{eqnarray}
&&B^0_s \to   {P}_{\bar{c}susu}^{0}  \overline \Lambda/\overline \Sigma^0, 
  B^+ \to    {P}_{\bar{c}susu}^{0}  \overline\Xi^+, 
  B^+\to   {P}_{\bar{c}sudu}^{(\prime) 0}  \overline \Sigma^+, 
  B^0_s \to   {P}_{\bar{c}sudu}^{(\prime) 0}  \overline n,   \nonumber\\
  &&
  B^0 \to   {P}_{\bar{c}sudu}^{(\prime) 0}  \overline \Lambda/ \overline \Sigma^0, 
  B^0 \to   {P}_{\bar{c}dudu}^{0}  \overline n, 
  B^0 \to   {P}_{\bar{c}dsdu}^{(\prime)-}  \overline\Sigma^+, 
  B^0\!\! \to\!\!   {P}_{\bar{c}dssu}^{(\prime)-}  \overline\Xi^+, 
  B^0_s\to   P_{\bar cdssu}^{(\prime)-}  \overline \Sigma^+.\label{eq:golden}
\end{eqnarray}}

We have taken into account both strong and weak decays of the {$P_{\bar{c}\boldsymbol{3}}$ and $P_{\bar{c}\boldsymbol{\bar 6}}$} states to prepare the reconstruction modes for experimental research, as shown in Tab.~\ref{tab:swdecay1}. For example {${P}_{\bar{c}sudu}^{(\prime)0}$} can be reconstructed via its strong decay mode {${P}_{\bar{c}sudu}^{(\prime) 0}\to D_{s}^{-}p$}, and {${P}_{\bar{c}susu}^{0}$} can be reconstructed via its weak decay channel {${P}_{\bar{c}susu}^{0}\to Kp$}. This result can help in optimizing experimental resources and detecting the {$P_{\bar{c}\boldsymbol{3}}$ and $P_{\bar{c}\boldsymbol{\bar 6}}$} states via their corresponding production channel.

It is expected that the production induced by the decay of $\bar{b}\to \bar{c}u\bar{d}/\bar{s}$ will be much greater in comparison to that caused by the decay of $b\to u\bar{c}d/s$. This is because the former decay channels, which involve the CKM element $V_{cb}V_{ud}$, are Cabibbo allowed. In contrast, the latter decay channels, which involve the CKM element $V_{ub}V_{cs}$, are Cabibbo-suppressed. Future research is anticipated to focus on the weak three-body decay process of the $B$ meson, which will provide further insights into the production and detection of {$P_{\bar{c}\boldsymbol{3}}$ and $P_{\bar{c}\boldsymbol{\bar 6}}$} for experimental reference.

\section{Appendix}\label{sec:appendix}
\subsection{The production of pentaquark states from the B decays}
In the above text, we have discussed the production of the pentaquark ground states from $B$ decays. Next, we will explore the production of the excited pentaquark states $\boldsymbol{15}$ and $\boldsymbol{15^{\prime}}$, which are denoted as $T_{\boldsymbol{15}}$ and $T_{\boldsymbol{15^{\prime}}}$ respectively. 
{In flavor space, the two diquarks of the excited pentaquark state $\boldsymbol{15}$ are mixing symmetry, ie. $\{[qq]_{\boldsymbol{\bar 3}}\{qq\}_{\boldsymbol{6}}\}_{\boldsymbol{15}}$ or $[\{qq\}_{\boldsymbol{6}}\{qq\}_{\boldsymbol{6}}]_{\boldsymbol{15}}$, while the ones of $\boldsymbol{15^{\prime}}$ are fully symmetry as $\{\{qq\}_{\boldsymbol{\bar 6}}\{qq\}_{\boldsymbol{\bar 6}}\}_{\boldsymbol{15}^{\prime}}$. 
}{According to the good diquark scheme~\cite{Jaffe:2004ph}, in color space the good diquark should be antisymmetry as $[qq]_{\boldsymbol{\bar 3}}(A)$, 
and in spin space the spin of antisymmetric diquark is $[qq]\to 0_{s}(A)$, While the ones of the symmetric diquark is $\{qq\}\to 1_{s}(S)$. So the spin of $T_{15}$ should be $1/2$ or $3/2$, and the ones of $T_{15^{\prime}}$ would be $3/2$ or $5/2$. 
Thus, under \textbf{flavor} $\boldsymbol{\otimes}$ \textbf{color} $\boldsymbol{\otimes}$ \textbf{spin} space, the wave function of the singly anti-charmed pentaquark states $T_{15}$ and $T_{15^\prime}$ can be written
  \begin{eqnarray}
  \psi_{\text{flavor}}=
  \left\{\begin{array}{c}\bar{c}\Big[\{qq\}_{\boldsymbol{6}}\{qq\}_{\boldsymbol{6}}\Big]_{15}\\ 
  \bar{c}\Big\{[qq]_{\boldsymbol{\bar{3}}}\{qq\}_{\boldsymbol{6}}\Big\}_{15}\\
  \bar{c}\Big\{\{qq\}_{\boldsymbol{6}}\{qq\}_{\boldsymbol{6}}\Big\}_{15^{\prime}}\end{array}\right. ,\ \
  \psi_{\text{color}}=\bar{c}_{\boldsymbol{\bar{3}}}\Big [[qq]_{\boldsymbol{\bar{3}}}[qq]_{\boldsymbol{\bar{3}}}\Big]_{3},\ \
  \psi_{\text{spin}}=\left\{\begin{array}{c}\bar{c}_{1\over 2}\Big[\{qq\}_{1}\{qq\}_{1}\Big]_{0}\\
  \bar{c}_{1\over 2}\Big\{[qq]_{0}\{qq\}_{1}\Big\}_{1}\\
  \bar{c}_{1\over 2}\Big\{\{qq\}_{1}\{qq\}_{1}\Big\}_{2}
  \end{array}\right. .
  \end{eqnarray} } Here the curly braces and square brackets indicate the symmetry (S) and antisymmetry (A) of the four light quarks respectively. The corresponding Hamiltonian for their production can be given as follows:
\begin{eqnarray}
  &&{\mathcal{H}}^{\prime}={\mathcal{H}}_{15'}+{\mathcal{H}}_{15}\nonumber\\
  &&=f_{1}B^{i}(H_{\boldsymbol{8}})^{j}_{l}(\overline{T}_{15'})_{\{ijkm\}}(P_{\bar8})_{\alpha}^{k} \varepsilon^{\alpha lm}+f_{2}B^{i}(H_{\boldsymbol{8}})^{j}_{i}(\overline{T}_{15})_{\{jl\}}^{k}(P_{\bar8})_{k}^{l} \nonumber\\
  &&+f_{3}B^{i}(H_{\boldsymbol{8}})^{j}_{k}(\overline{T}_{15})_{\{il\}}^{k}(P_{\bar8})_{j}^{l}
  +f_{4}B^{i}(H_{\boldsymbol{8}})^{j}_{l}(\overline{T}_{15})_{\{ij\}}^{k}(P_{\bar8})_{k}^{l}
  +f_{5}B^{i}(H_{\boldsymbol{8}})^{j}_{k}(\overline{T}_{15})_{\{jl\}}^{k}(P_{\bar8})_{i}^{l}\nonumber\\
  &&+\bar{f}_{1} \overline{B}_{i}(H_{\boldsymbol{6}})^{\{jk\}}(\overline{T}_{15'})_{\{jklm\}}(P_{\bar8})_{\alpha}^{l} \varepsilon^{\alpha im}+\bar{f}_{2} \overline{B}_{i}(H_{\boldsymbol{\bar{3}}})^{[ij]}(\overline{T}_{15})_{\{jk\}}^{l}(P_{8})_{l}^{k}
     +\bar{f}_{3} \overline{B}_{i}(H_{\boldsymbol{\bar{3}}})^{[lj]}(\overline{T}_{15})_{\{jk\}}^{i}(P_{\bar8})_{l}^{k}\nonumber\\
&&
     +\bar{f}_{4}\overline{B}_{i}(H_{\boldsymbol{6}})^{\{ij\}}(\overline{T}_{15})_{\{jk\}}^{l}(P_{\bar8})_{l}^{k}
     +\bar{f}_{5}\overline{B}_{i}(H_{\boldsymbol{6}})^{\{jl\}}(\overline{T}_{15})_{\{jk\}}^{i}(P_{\bar8})_{l}^{k}
     +\bar{f}_{6}\overline{B}_{i}(H_{\boldsymbol{6}})^{\{jk\}}(\overline{T}_{15})_{\{jk\}}^{l}(P_{\bar8})_{l}^{i}.\label{eq:hamiltonian615p}
\end{eqnarray}
For simplicity, we present a useful reference for reconstructing the strong decays of the pentaquark ground states by arranging the decay width relation as follows:
{
\begin{eqnarray}
   && \Gamma(P_{\bar csudu}^{0}\to \overline D^0 \Lambda^0)
   = \frac{1}{3}\Gamma(P_{\bar csudu}^{0}\to \overline D^0 \Sigma^0)
   = \frac{1}{6}\Gamma(P_{\bar csudu}^{0}\to D^- \Sigma^+)\nonumber\\
   &&= \frac{1}{6}\Gamma(P_{\bar csudu}^{0}\to  D^-_s p)
   = \frac{1}{6}\Gamma(P_{\bar cdsdu}^{-}\to \overline D^0 \Sigma^-)
   = \Gamma(P_{\bar cdsdu}^{-}\to D^- \Lambda^0)\nonumber\\
   &&= \frac{1}{3}\Gamma(P_{\bar cdsdu}^{-}\to D^- \Sigma^0)
   = \frac{1}{6}\Gamma(P_{\bar cdsdu}^{-}\to  D^-_s n)
   = \frac{1}{6}\Gamma(P_{\bar cdssu}^{-}\to \overline D^0 \Xi^-)\nonumber\\
   &&= \frac{1}{6}\Gamma(P_{\bar cdssu}^{-}\to D^- \Xi^0)
   = \frac{1}{4}\Gamma(P_{\bar cdssu}^{-}\to  D^-_s \Lambda^0),\label{eq:strongre3}\\ 
&&\Gamma({P}_{\bar{c}dudu}^{0}\!\to\! \overline D^0 n)= \Gamma({P}_{\bar{c}dudu}^{0}\!\to\! D^- p)= \Gamma(P_{\bar{c}dsds}^{--}\!\to\! D^- \Xi^-)= \Gamma(P_{\bar{c}dsds}^{--}\!\to\!  D^-_s \Sigma^-)\nonumber\\
&&= \frac{2}{3}\Gamma({P}_{\bar{c}sudu}^{\prime 0}\!\to\! \overline D^0 \Lambda)= 2\Gamma({P}_{\bar{c}sudu}^{\prime 0}\!\to\! \overline D^0 \Sigma^0)
= \Gamma({P}_{\bar{c}sudu}^{\prime 0}\!\to\! D^- \Sigma^+)=\Gamma({P}_{\bar{c}sudu}^{\prime 0}\!\to\!  D^-_s p)\nonumber\\
&&= \Gamma({P}_{\bar{c}dsdu}^{\prime-}\!\to\! \overline D^0 \Sigma^-)
= \frac{2}{3}\Gamma({P}_{\bar{c}dsdu}^{\prime-}\!\to\! D^- \Lambda)= 2\Gamma({P}_{\bar{c}dsdu}^{\prime-}\!\to\! D^- \Sigma^0)
= \Gamma({P}_{\bar{c}dsdu}^{\prime-}\!\to\!  D^-_s n)\nonumber\\
&&= \Gamma({P}_{\bar{c}susu}^{0}\!\to\! \overline D^0 \Xi^0)= \Gamma({P}_{\bar{c}susu}^{0}\!\to\!  D^-_s \Sigma^+)
= \Gamma({P}_{\bar{c}dssu}^{\prime-}\!\to\! \overline D^0 \Xi^-)\nonumber\\
&&= \Gamma({P}_{\bar{c}dssu}^{\prime-}\!\to\! D^- \Xi^0)
= \frac{1}{2}\Gamma({P}_{\bar{c}dssu}^{\prime-}\!\to\!  D^-_s \Sigma^0).\label{eq:strongre6}
  \end{eqnarray}}
\section*{Acknowledgements}
This work is supported in part by the youth Foundation JN210003, of China University of mining and technology.

\textbf{Data Availability Statement:}  
No Data is associated with the manuscript.

\end{document}